\documentclass[useAMS,usenatbib]{mn2e}
\usepackage{graphicx,amssym}
\newcommand{\bc}{\begin{center}}
\newcommand{\ec}{\end{center}}

\title[Modelling the SED dependence of galaxy clustering ]
      {Modelling and interpreting 
        the dependence of  clustering on the spectral energy 
	distributions of galaxies}
\author[L.Wang, C.Li, G.Kauffmann, G.~De Lucia]
       {Lan Wang$^{1,}$$^2$$\thanks{Email: wanglan@mpa-garching.mpg.de}$,
	Cheng Li$^{2,3,4,5}$,
        Guinevere Kauffmann$^2$,
	Gabriella De Lucia$^2$   
        \\      
	$^1$Department of Astronomy, Peking University, Beijing 100871, China\\
        $^2$Max--Planck--Institut f\"ur Astrophysik, 
        Karl--Schwarzschild--Str. 1, D-85748 Garching, Germany\\
	$^3$The Partner Group of MPI f\"ur Astrophysik,
      Shanghai Astronomical Observatory,
      Nandan Road 80, Shanghai 200030, China\\
      	$^4$Center for Astrophysics, University of Science
      and Technology of China, Hefei, Anhui 230026, China\\
	$^5$Joint Institute for Galaxy and Cosmology (JOINGC) of SHAO and USTC}
\begin{document}

\date{Accepted 2007 ???? ??. 
      Received 2007 ???? ??; 
      in original form 2007 ???? ??}

\pagerange{\pageref{firstpage}--\pageref{lastpage}} 
\pubyear{2007}

\maketitle

\label{firstpage}

\begin{abstract}
We extend our previous physically-based halo occupation distribution models  
to include the dependence of  clustering on the spectral
energy distributions  of galaxies. The high resolution 
{\it Millennium Simulation} is used to specify the                          
positions and the velocities of the model  galaxies. The stellar mass
of a galaxy is assumed to depend only on $M_{infall}$, the halo mass when
the galaxy was last the central dominant object of its halo.
Star formation histories are parametrized using 
two additional quantities that are measured from the simulation for each galaxy:
its  formation time ($t_{form}$), and the time when it 
first  becomes a satellite ($t_{infall}$).
Central galaxies begin forming stars at time $t_{form}$  
with an exponential  time scale $\tau_c$. 
If the galaxy becomes a satellite, its star formation  
declines thereafter with a new time scale $\tau_s$. 
We compute  4000 \AA\ break strengths for our model galaxies using 
stellar population synthesis models.
By fitting these models to the  observed abundances  and projected correlations of galaxies   
as a function of  break strength in  the  Sloan
Digital Sky Survey, we  constrain $\tau_c$ and $\tau_s$  
as  functions of galaxy stellar mass.  
We find that central galaxies  with large stellar masses
have ceased forming stars. At low stellar masses, central galaxies display
a wide range of different star formation histories, with a significant 
fraction experiencing recent starbursts. Satellite galaxies of all masses 
have declining star formation rates, with similar e--folding times,    
$\tau_s \sim  2.5$ Gyr. One consequence of this 
long e--folding time is that the colour--density relation is predicted to
flatten at  redshifts  $> 1.5$, because  star formation in 
the majority of satellites has
not yet declined  by a significant factor. This is consistent  
with recent observational results from the DEEP and VVDS surveys.
\end{abstract}


\section{Introduction}
\label{sec:intro}

The most fundamental metric of a galaxy is its luminosity, which
serves as a rough indicator of its total mass.
Another fundamental metric of a galaxy is its  colour
or spectral type, which is usually interpreted as an indicator of its
recent star formation history (although
metallicity and dust are both known to affect galaxy colours). 
In the local Universe, galaxy luminosities and colours 
are known to be strongly correlated; luminous elliptical galaxies are much 
redder than the less luminous spirals. It has also long been known that 
the clustering of galaxies is a strong function
of morphological type \citep{hubble1936,abell1958,davis1976,dressler1980,
loveday1995,zehavi2002}, spectral type \citep{loveday1999,norberg2002,
madgwick2003}, and colour \citep{willmer1998,zehavi2002,budavari2003}. 
On the other hand, galaxy clustering also            
depends on luminosity \citep{hamilton1988,park1994,willmer1998,zehavi2002}, 
with luminous galaxies being more strongly clustered than fainter ones.

Recent large surveys such as 
2dfGRS \citep{colless2001} and SDSS \citep{york2000} have allowed the
{\em covariance} between galaxy luminosity and colour to be broken.
\citet{norberg2002} showed at all luminosities, galaxies  with spectral 
features indicative of a ``passive'' old  stellar population
have higher correlation amplitude than galaxies with ongoing star formation.
Likewise \citet{zehavi2005} and \citet{li2006a}  measured the clustering
of red and blue galaxies as a function of luminosity and stellar mass, and 
showed that red galaxies are more strongly correlated and have a correlation 
function with a steeper slope. 
The strongest  differences between the
red and blue correlation functions occurred for galaxies with
the lowest luminosities and stellar masses.

In order to interpret these results, we need to understand the underlying
physics that causes the dependence of the correlation function on
colour/spectral type. Semi-analytic models of galaxy formation are able to
provide considerable insight into the processes that determine how
galaxies with different properties cluster  
\citep{kauffmann1997,kauffmann1999,benson2000,croton2005}. These models use    
N-body simulations of the dark matter to specify the
locations of galaxies, and they invoke simple prescriptions to describe  
processes such as 
gas cooling, star formation and supernova feedback. 

In the semi-analytic models (SAMs), there are two reasons why galaxies 
transition from star-forming (blue) to passive (red) systems. One is 
a consequence of the infall of a galaxy onto a larger halo.
When this occurs, the galaxy is stripped of its supply of infalling gas 
and its star formation rate  declines as its cold gas reservoir is depleted. 
This means that there will be a population of red {\em satellite} galaxies
located in groups and clusters \citep{diaferio2001}.
However, this process by itself  is not sufficient to explain the very strong observed 
dependence of colour on galaxy luminosity. Some other process must
act to terminate star formation in the {\em central} galaxies of massive
dark matter haloes. In recent models \citep{croton2005, bower2006,
cattaneo2006b}, feedback from active galactic nuclei (AGN) has been 
invoked as a possible mechanism for suppressing star formation in massive
central galaxies. However, the details of the AGN feedback process and
the truncation of star formation in infalling satellite galaxies
remain poorly constrained, so in reality there is still considerable
freedom when attempting to specify  the star formation histories
of galaxies in these models. The colour distributions
generated by  the  SAMs should thus be treated as
indicative rather than quantitative predictions of the models.

The Halo Occupation Distribution (HOD) approach bypasses
any consideration of the physical processes important in
galaxy formation. It specifies how
galaxies are related to dark matter halos in a purely statistical fashion.
\citet{bosch2003} was the first to model the clustering properties
of early and late-type galaxies in the context of HOD models 
by using observational data from the 2dF redshift survey to constrain the
average number of early and late-type galaxies of given luminosity
residing in a halo of given mass.
More recently, \citet{zehavi2005} built HOD models
that were able to reproduce the correlation function of red and blue galaxies as a 
function of luminosity, by defining a  blue galaxy fraction as a 
function of dark halo mass. The blue fraction also depended on whether the
galaxy was  a central or a satellite system. 
The results of both studies demonstrate that the strong clustering 
of faint red galaxies can be explained if nearly all of them 
are satellite systems in high-mass halos. These results have
recently been extended to higher redshifts by \citet{phleps2006}. 
The main  disadvantage
of the HOD approach is that it remains unclear how one progresses from
a purely statistical characterization of the link between galaxies
and dark matter halos, to a more physical understanding of
the galaxy formation process itself. 

In previous work \citep{wang2006}, we attempted to build a  
{\em physically-based} HOD model that would combine  the advantages of the
statistical HOD models with those of the semi-analytic approach. A large high
resolution N-body simulation, the {\it Millennium Simulation} 
\citep{springel2005} was used to follow the merging paths of dark matter
halos and their associated substructures and to specify the positions and
the velocities of the galaxies in the simulation box.   
The properties of the galaxies (in this case, their
luminosities and stellar masses) 
were specified using simple parameterized
functions.  \citet{wang2006} chose to parametrize the luminosities and masses
of the galaxies in their models in terms of the quantity $M_{infall}$, which
was defined as the mass of the dark matter halo when the galaxy was last the
central galaxy of its own halo. \citet{wang2006} tested this parametrization
using the semi-analytic galaxy catalogues of \citet{croton2005} and and then applied
it to data from the Sloan Digital Sky Survey. They were able to show that
the relation between stellar mass and halo mass inferred from the combination
of the models and the clustering data was in good agreement with independent
measurements of this relation using galaxy-galaxy lensing techniques 
\citep{mandelbaum2006}.

In this paper, we extend our method to model the 
dependence of clustering on the spectral energy distributions of galaxies.  
For galaxies of given stellar mass, we assume the star formation history to
depend not only on stellar mass, but also on whether the galaxy is a central
object or a satellite in the simulation. We fit this model to the 
 colour distributions of galaxies 
of give  stellar mass,  as well as to their  correlation functions split by colour.
We choose to  focus on the spectral 
index D$4000$ rather than a  more traditional broad-band colour. 
The D$4000$ index is defined as the ratio of the flux in two bands at the 
long-- and short--wavelength side of the $4000$\AA \hspace{0.1cm} 
discontinuity. This $4000$\AA \hspace{0.1cm} break arises from a sum of many 
absorption lines produced by ionized metals in the atmosphere of stars.
Because the absorption increases with decreasing stellar temperature, 
the D$4000$
break gets larger with older ages, and is largest for old and metal--rich
stellar populations. Therefore it is a good indicator of the star formation 
history of a galaxy. In this work we adopt the narrow definition of the 
$4000$\AA \hspace{0.1cm} break 
introduced by \citet{balogh1999}, and denote it as D$_n4000$.
Galaxies with large and small values of D$_n4000$ are referred to as 
``old'' and ``young'' respectively.
Because D$_n4000$ is defined in a narrow wavelength 
interval, it is  insensitive to the effects of dust. 

The parametrized  models that we construct extend the old ones in which
stellar mass $M_*$ is assumed to depend only on $M_{infall}$ by assuming that
the star formation history of each galaxy declines exponentially after its first appearance
in the simulation with time constant $\tau_c(M_*)$ as long as it is a central galaxy, and then
with a different time constant $\tau_s(M_*)$ after it becomes a satellite galaxy.   
Note that this strongly resembles the approach adopted in the semi-analytic models.
The main difference is that the time scales $\tau_c$ and $\tau_s$
and their dependence on mass are not specified using any fixed 
``recipe''; we allow these time scales to be directly constrained 
by the observational data. As we will see, this approach allows us to draw
interesting physical conclusions from  a comparison of our models with the SDSS data.

In Sec.~\ref{sec:observation} we present observational results from
the Sloan Digital Sky Survey (SDSS). These include the 
D$_n4000$/colour distributions of galaxies in different stellar mass
ranges and the projected two point correlation function of
red/blue and old/young galaxies.
In Sec.~\ref{sec:concepts}, we outline our basic theoretical concepts 
and describe how we decided to model the star formation histories
of the galaxies, and calculate their spectral energy distributions. In 
Sec.~\ref{sec:fitting} we describe how we fit the models to the data.  
Interpretation and tests of our 
results are given in Sec.~\ref{sec:tests}. Conclusions
and discussions are presented in the final section.


\section{The observational results}
\label{sec:observation}

Our observational results are based on a sample of $\sim$200,000 galaxies 
drawn from the SDSS Data Release Two. The galaxies have 
$0.01 < z <  0.3$, $14.5<r<17.77$ 
and $-23< M_{^{0.1}r}<-16$, where $r$ is the $r$-band Petrosian apparent 
magnitude corrected for foreground extinction, and $M_{^{0.1}r}$ 
is the $r$-band absolute magnitude corrected to redshift $z=0.1$. 
This sample has formed the basis of our previous investigations of 
the correlation function, the power spectrum, pairwise velocity dispersion
distributions, and the
luminosity and the stellar mass function of galaxies \citep{li2006a,li2006b}. 
In \citet{wang2006}, we made use of the measurements of the projected 
correlation function for galaxies in bins of luminosity and stellar mass 
to constrain the relation between galaxy luminosity/stellar mass and 
$M_{infall}$. In this paper, we extend this analysis to the  
spectral energy distributions of galaxies. In this
section, we  focus on the 
distributions of  $D_n4000$ and colour and we explore how thee projected two-point correlation
functions split by  $D_n4000)$ and colour 
change as a function of $M_*$.

\begin{figure*}
\bc
\hspace{-1.6cm}
\resizebox{17.cm}{!}{\includegraphics{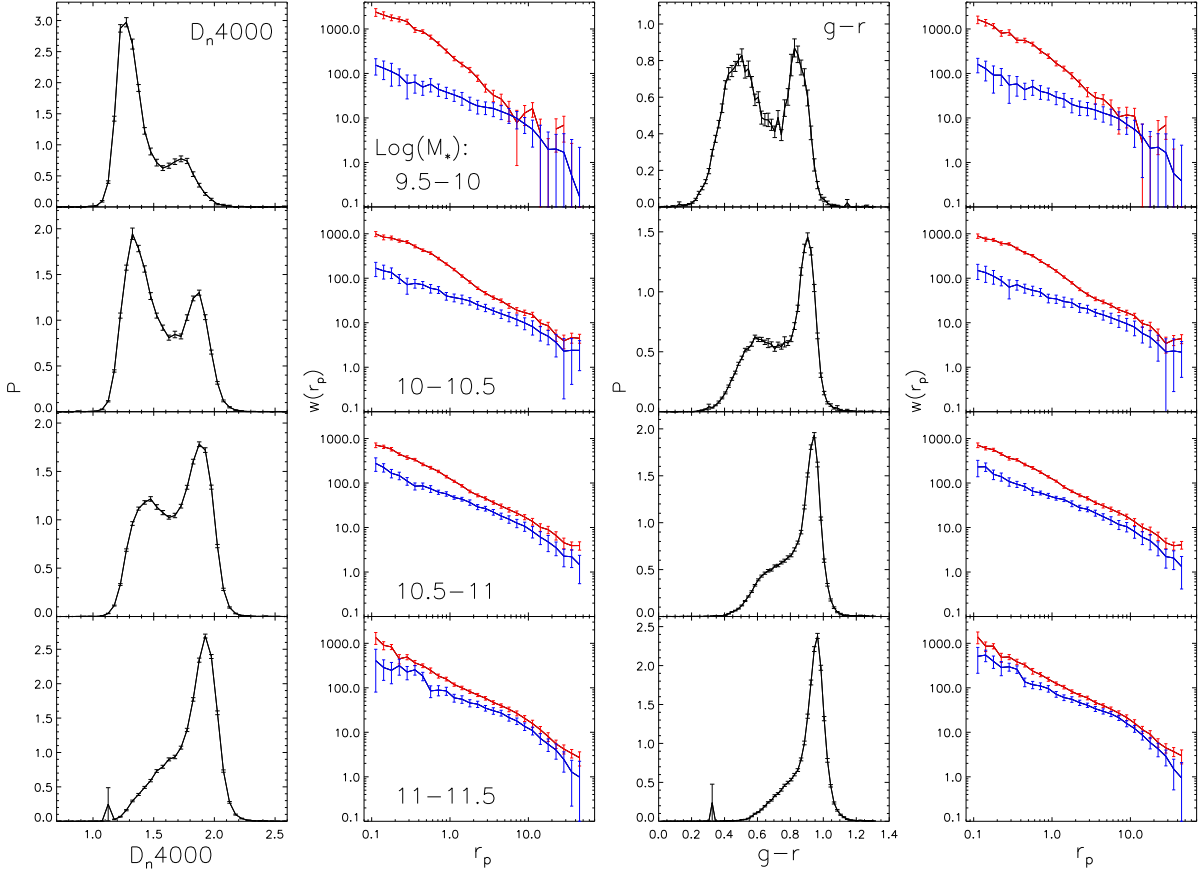}}\\%
\caption{  
Observational results from SDSS in bins of stellar mass. The  left two 
columns  show  D$_n4000$ distributions and correlation functions split 
by D$_n4000$. Red/blue lines represent subsamples with larger/smaller 
values of D$_n4000$. The right two columns  are for $g-r$; the 
red and blue lines here represent clustering of red and blue subsamples.
}
\label{fig:SDSS}
\ec
\end{figure*}

\begin{figure*}
\bc
\hspace{-1.6cm}
\resizebox{17.cm}{!}{\includegraphics{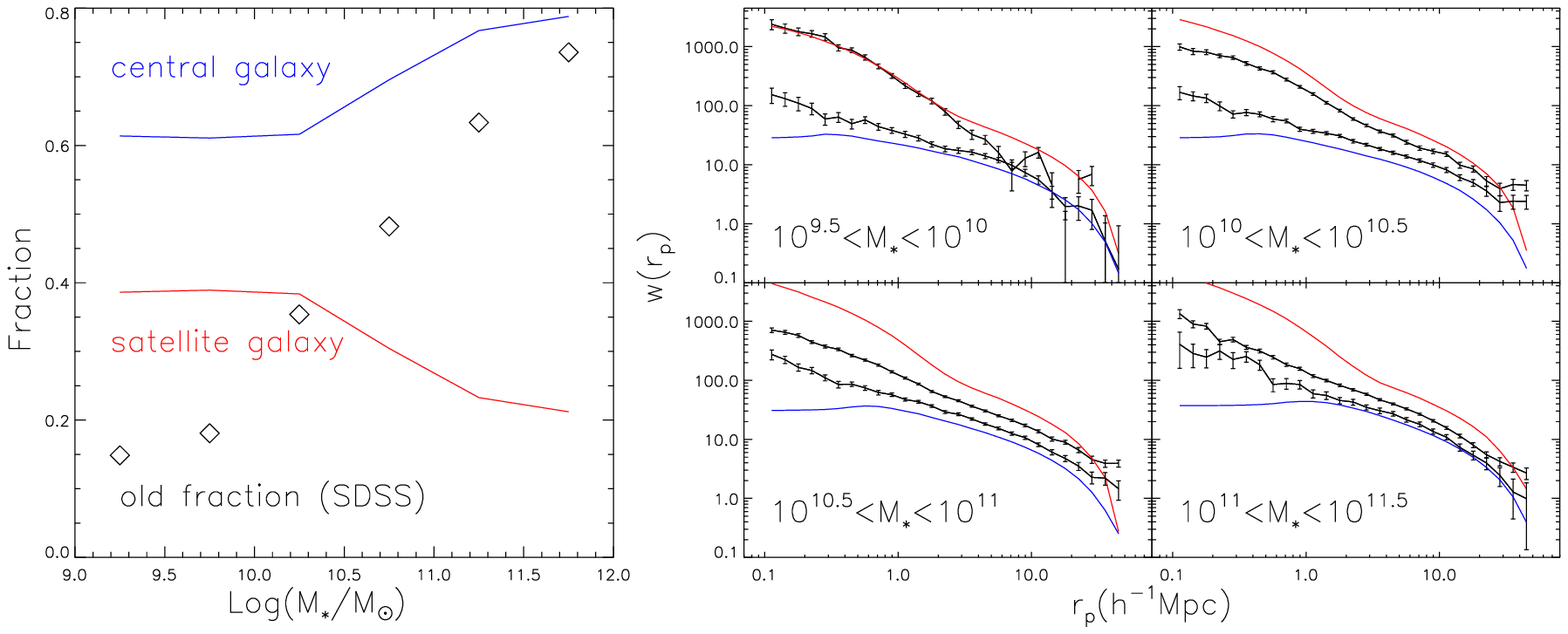}}\\%
\caption{  
Left panel: fractions of central (blue) and satellite (red)
   systems in bins of stellar mass in the models of                         
\citet{wang2006}. Black symbols show  the fraction of
galaxies in the high D$_n4000$ peak from  SDSS.
Right panel: the  clustering of central (blue) and 
satellite (red) galaxies in the model  compared with low/high  
D$_n4000$ subsamples in the  SDSS (upper and lower black curves ).
}
\label{fig:fraction}
\ec
\end{figure*}

The galaxies in our sample are divided into four subsamples according to 
their stellar mass. Each of the stellar mass subsamples is then divided into 
two 
further subsamples according to D$_n4000$ and $g-r$, using a method similar 
to that adopted in \citet{li2006a}.  
We  fit bi-Gaussian 
functions to the distribution of D$_n4000$ and $g-r$ as a function of stellar mass.   
The division  into high D$_n4000$ 
and low D$_n4000$, red and blue, is defined to be  the mean of the 
two Gaussian centres in each stellar mass bin. In the computation of the 
D$_n4000$ and $g-r$ distributions, we have corrected for incompleteness by
weighting each galaxy a factor of $V_{survey}/V_{max}$. $V_{survey}$ 
is the volume for the sample, and $V_{max}$ is the maximum volume over which 
the galaxy could be observed within the sample redshift range and within the
range of $r$-band apparent magnitudes.

For each subsample, the redshift-space two point correlation function(2PCF) 
$\xi^{s}(r_p,\pi)$ is
measured using the \citet{hamilton1993} estimator.
The projected 2PCF $w(r_p)$ is then estimated by
integrating $\xi^{s}(r_p,\pi)$ along the line-of-sight direction $\pi$,
with $|\pi|$ ranging from 0 to 40 $h^{-1}$ Mpc.
When computing 2PCFs, two different methods for constructing random
samples have been used: the ``standard'' method in which the redshift
selection function is explicitly modelled using  the luminosity function,
and the more ``general'' method in which only the sky positions of the
observed galaxies are randomly re-assigned.
As shown in \citet{li2006a}, the 2PCFs obtained with
the two methods are in good agreement, and here we will use the
more ``general'' method. We have also carefully corrected for possible biases, 
such as the variance in mass-to-light ratio and the small-scale deficiency 
in the 2PCF due to fibre collisions \citep{li2006a}. This ensures accurate 
measurements for correlation functions on small scales.

To take into account the effect of ``cosmic variance'' on the $w(r_p)$ 
measurements, we have constructed a set of 10 mock galaxy catalogues from the 
Millennium simulation with exactly the same geometry and 
selection function
as  the real sample. The effect of cosmic variance is modelled by placing 
a virtual observer randomly inside the simulation box when constructing these 
mock catalogues. The detailed procedure for constructing these mock 
catalogues has been presented in \citet{li2006c}.
For each mock catalogue, we divide galaxies into subsamples according to
stellar mass and $g-r$ colour, in the same way as for the real sample, and we 
measure $w(r_p)$ for these subsamples. The $1\sigma$ variation between 
these mock catalogues is then added in quadrature to the 
bootstrap errors. Note that the errors are assumed to be the same for the 
splits by $g-r$ and by  D$_n4000$.  

Fig.~1 shows the distributions of D$_n4000$ 
and $g-r$ colour for galaxies, as well as the projected 2PCF 
$w(r_p)$ for the ``red'' and ``blue'' subsamples in four stellar mass ranges.
As can be seen, in each stellar mass interval, D$_n4000$/$g-r$ shows a bimodal
distribution, with the fraction of galaxies in the red peak increasing 
towards higher masses. Older/redder galaxies of all stellar 
masses are more strongly clustered and have steeper correlation functions 
than their younger/bluer counterparts. This age/colour dependence is much 
stronger for the low mass galaxies than for the high mass galaxies, 
particularly on small scales. Notice that the clustering for subsamples split 
by $D_n4000$ and $g-r$ are quite similar, but   
the distribution functions of these two quantities are quite different, 
particularly at low masses.
For example, for the lowest mass bin ($9.5<\log(M_{*}/M_\odot)<10$),
the fractions of red and blue galaxies are comparable, but the
fraction of galaxies with large D$_n4000$ values is much smaller than that of 
the low D$_n4000$ population.  
When building our model, we will first focus on the
D$_n4000$ spectral index, and test to what extent a model that reproduces
the observed trends as a function of D$_n4000$  will
also work for $g-r$ colour.

\section{theoretical concepts}
\label{sec:concepts}

In the paper of \citet{wang2006}, we used the {\it Millennium Simulation}
to construct a model to describe the clustering of
galaxies as a function of their luminosities and
 stellar masses. The positions 
and velocities of the galaxies in the simulation box
were obtained by following the orbits and merging 
histories of the substructures in the simulation. Parametrized functions 
were then adopted to relate the luminosities and stellar masses
of the galaxies to the quantity 
$M_{infall}$, defined as the mass of the halo at the epoch when the galaxy 
was last the central dominant object in its own halo. By fitting both the stellar 
mass function and the projected
correlation function $w(r_p)$ measured in five different 
stellar mass bins, we were able to use the SDSS data to constrain the link between
galaxy stellar mass $M_{*}$  and $M_{infall}$. 
This relation  was parametrized using a 
double power-law of the form:
\begin{displaymath}
{{M}_{*}} = \frac{2}{(\frac{{M}_{infall}}{{{M}_{0}}})^{-\alpha}+(\frac{{ M}_{infall}}{{{M}_{0}}})^{-\beta}}{\times}{k},
\end{displaymath}
The scatter in  $\log(M_{*})$ at a given value
of $M_{infall}$  was  described using a Gaussian function with 
width $\sigma$. 
Our best-fit model to the SDSS data had the following parameters: $M_0=4.0\times10^{11}h^{-1}M_{\odot}$, 
$\alpha=0.29$, $\beta=2.42$, $\log{k}=10.35$ and $\sigma=0.203$ for  
central galaxies and $M_0=4.32\times10^{11}h^{-1}M_{\odot}$, $\alpha=0.232$, 
$\beta=2.49$, $\log{k}=10.24$ and $\sigma=0.291$ for satellite galaxies.
{\footnote {Note these parameters are slightly
different from those  published in the paper of \citet{wang2006},
because there was  a small change in the definition of ``first progenitor''
when  building the dark matter subhalo tree in the simulation (see  
\citet{lucia2006} for more details about this modification). Nevertheless, 
the best fit relation is almost the same as previously obtained, and all the 
conclusions of the paper remain unchanged.}}

In this paper, our aim is to extend this model to describe not only the
masses and luminosities of galaxies, but also their colours and spectral
energy distributions. A successful model should be able to reproduce the
 4000 \AA\ break strength  distribution at each stellar mass, as well as the 
correlation functions split by D$_n$(4000) shown in Fig. 1.   

In standard semi-analytic models, galaxies that reside at the centre of their 
dark matter halo are called central galaxies, 
and those that have been accreted into larger structures
are termed satellite galaxies. The simplest model one could imagine 
for differentiating galaxies in the  high/low D$_n4000$  or
red/blue peaks of the bimodal distributions shown in Fig. 1
would be that these two peaks correspond to these
satellite and central galaxy populations.
Central galaxies have young stellar populations and  
more active star formation because of ongoing cooling and gas accretion. 
Satellite galaxies are older because they run out of gas after 
they are accreted or their gas is removed by processes such as ram pressure 
stripping. In early  semi-analytic models \citep{kauffmann1993,
cole1994, somerville1999}, this was indeed the case;
red galaxies were mainly satellite galaxies and blue 
galaxies were the  central galaxies of their own haloes.

\begin{figure}
\bc
\hspace{-0.6cm}
\resizebox{8.5cm}{!}{\includegraphics{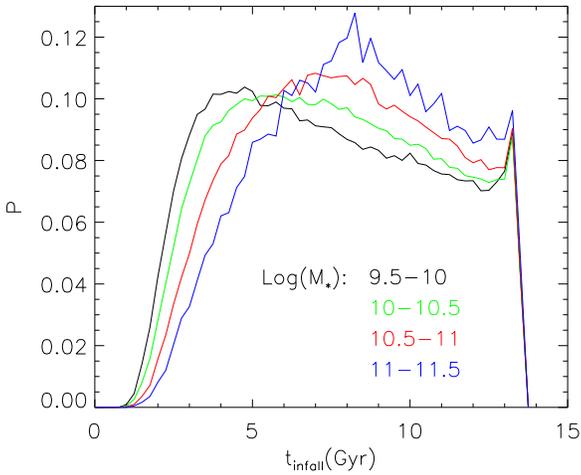}}\\%
\caption{Distribution
of infall times for satellite galaxies in different stellar 
mass bins. The x-axis shows time from past to present, with 13.7Gyr 
corresponding to the present day.
}
\label{fig:accretion}
\ec
\end{figure}

Fig.~\ref{fig:fraction} shows that this picture does not fit the SDSS 
observations.
In the left panel of Fig.~\ref{fig:fraction}, blue and red lines show        
the fraction of central and satellite  galaxies as a function of stellar
mass in the model of \citet{wang2006}.  
For comparison,  diamonds show the fraction of galaxies in the high D$_n4000$
peak as a function of $M_*$, as measured  from the SDSS data. 
As can be seen, the fraction of satellite galaxies in the simulation
does not match the fraction of ``old'' galaxies in the real
Universe, except at stellar masses $\sim 10^{10} M_{\odot}$. At lower
stellar masses, the fraction of satellite galaxies is {\em higher}
than the fraction of old galaxies, implying that some low mass
satellite systems are still forming stars. At high stellar masses,
nearly all galaxies in the real Universe are ``old". Fig.~\ref{fig:fraction}  
shows
that the majority of these massive old galaxies must be central galaxies.
 
In the right panel of Fig.~\ref{fig:fraction},
blue and red lines show the projected two-point correlation
functions  for central and satellite galaxies in the simulations  in 4  different 
stellar mass ranges. Results from the SDSS for the low and high D$_n4000$    
subsamples are plotted in black. This again shows the
failure of a simple central/satellite dichotomy  as a way of explaining
the difference between the  ``old'' and ``young'' galaxy
populations in the SDSS. As can be seen,
the difference in clustering amplitude between the two  galaxy populations
{\em decreases} as a function of increasing stellar mass, while the
difference in clustering strength between central galaxies
and satellite galaxies remains approximately constant.   

It is important to remember that satellite galaxies were not all ``created''
at the same time.
Fig.~\ref{fig:accretion} shows distributions of the times at which satellite
galaxies of different stellar masses were first accreted by a larger 
structure. The satellite infall(i.e. accretion) times $t_{infall}$ 
are randomly distributed  between the two simulation snapshots when the 
galaxy first transitions from a being central object  
in its own halo to a satellite system. As can be seen,
high  mass satellites have on average been  accreted more recently than low 
mass satellites. This effect  goes in the {\em wrong direction} 
to resolve the discrepancies shown in Fig.~\ref{fig:fraction}. 
As we have discussed, a substantial number of low
mass satellite galaxies are required to have  ``young'' stellar
populations, but  Fig.~\ref{fig:accretion} shows that
low mass satellites  typically become satellites quite early on.

\begin{figure*}
\bc
\hspace{-1.6cm}
\resizebox{17.cm}{!}{\includegraphics{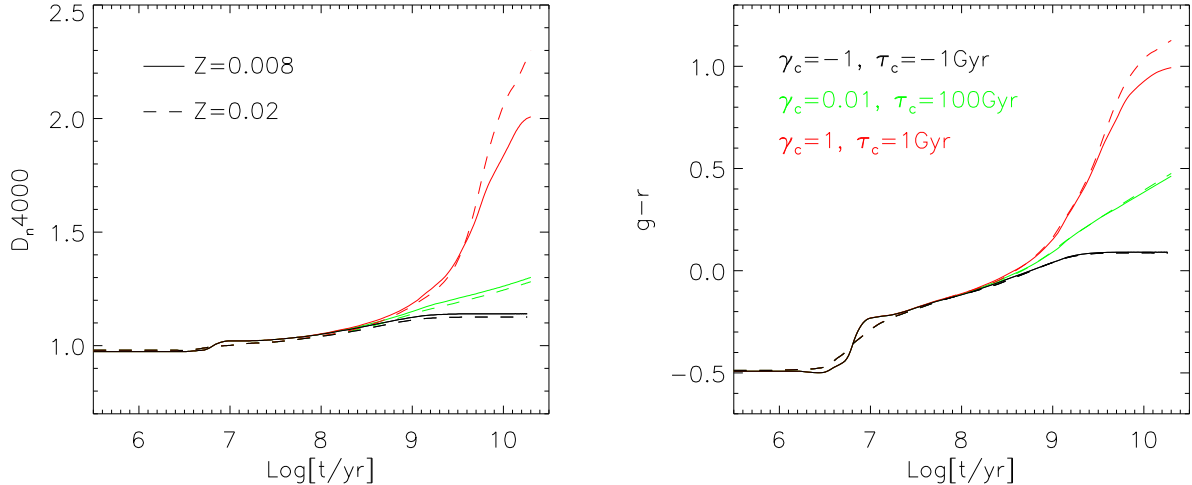}}\\%
\caption{  
The evolution of  
D$_n4000$/$g-r$ as a function of time for different values
of the star formation timescale  parameter $\tau_c$ for typical central 
galaxies. Solid lines are for solar metallicity  and dashed lines show
results for 0.25 solar metallicity. The coloured lines show results for three
different values of $\tau_c$.
}
\label{fig:BC03}
\ec
\end{figure*}

\begin{figure*}
\bc
\hspace{-1.6cm}
\resizebox{17.cm}{!}{\includegraphics{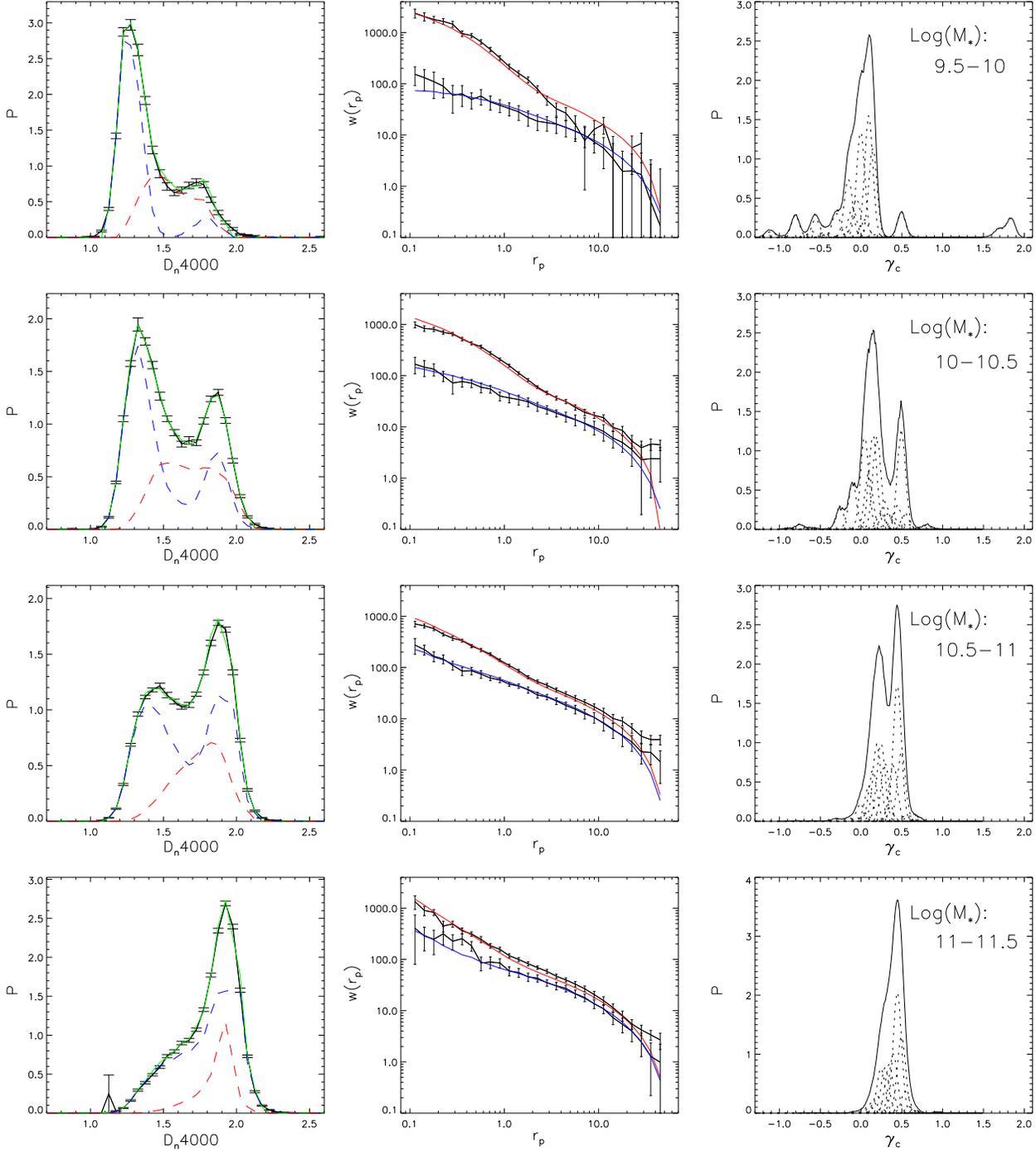}}\\%
\caption{  
Our best fits to the  D$_n4000$ distributions and the correlation functions
split by D$_n4000$ in different stellar mass bins. In the left panels, 
green lines are the full fits, while red/blue dashed lines show the
contributions from satellite/central galaxies. In the middle panels, 
red/blue lines are best-fit  correlation functions for subsamples with 
larger/smaller value of D$_n4000$. The   SDSS results are shown in black.
The right panels show the       
the distributions of $\gamma_{c}$ recovered by our
non-parametric technique, with dashed lines showing
the contribution of each individual Gaussian.
}
\label{fig:bestD4k}
\ec
\end{figure*}

 \begin{table*}
 \caption{Best--fit parameter values for  star formation 
histories in four stellar mass bins as derived from SDSS D$_n4000$ 
distribution and galaxy correlation functions split by D$_n4000$.}
\begin{center}
 \begin{tabular}{cccccccccc} \hline
   $log(M_{*}/M_{\odot})$ & $\gamma_k$ & $k$ & $<\gamma_c>_{median}$ & $\tau_{c}(Gyr)(50\%, 16\%, 84\%)$ & $<\tau_{s}>(Gyr)$ & $\sigma_{\tau_{s}}$ & ${\Xi}$ & $\chi^2_{dis}/N_{dis}$\\ \hline
9.5-10  &[-1.6,2.2] & 77   & 0.025 & (39.97, -4.589, 5.767) & 2.312 & 0.992& 4.214  & 3.126\\
10-10.5 &[-1,1.2]   & 45   & 0.175 & (5.724, 69.95, 2.106) & 2.364 & 1.041& 4.014 & 2.639\\
10.5-11 & [-1,1]    & 41  & 0.318 & (3.147, 6.900, 2.081) & 2.491 & 0.534& 5.267  & 4.239\\
11-11.5 & [-0.8,1.2]& 41  & 0.411 & (2.435, 3.845, 1.953) & 2.103 & 0.082& 7.061 & 5.837\\ \hline
 \end{tabular}
\end{center}
 \end{table*}

\begin{figure*}
\bc
\hspace{-1.6cm}
\resizebox{17.cm}{!}{\includegraphics{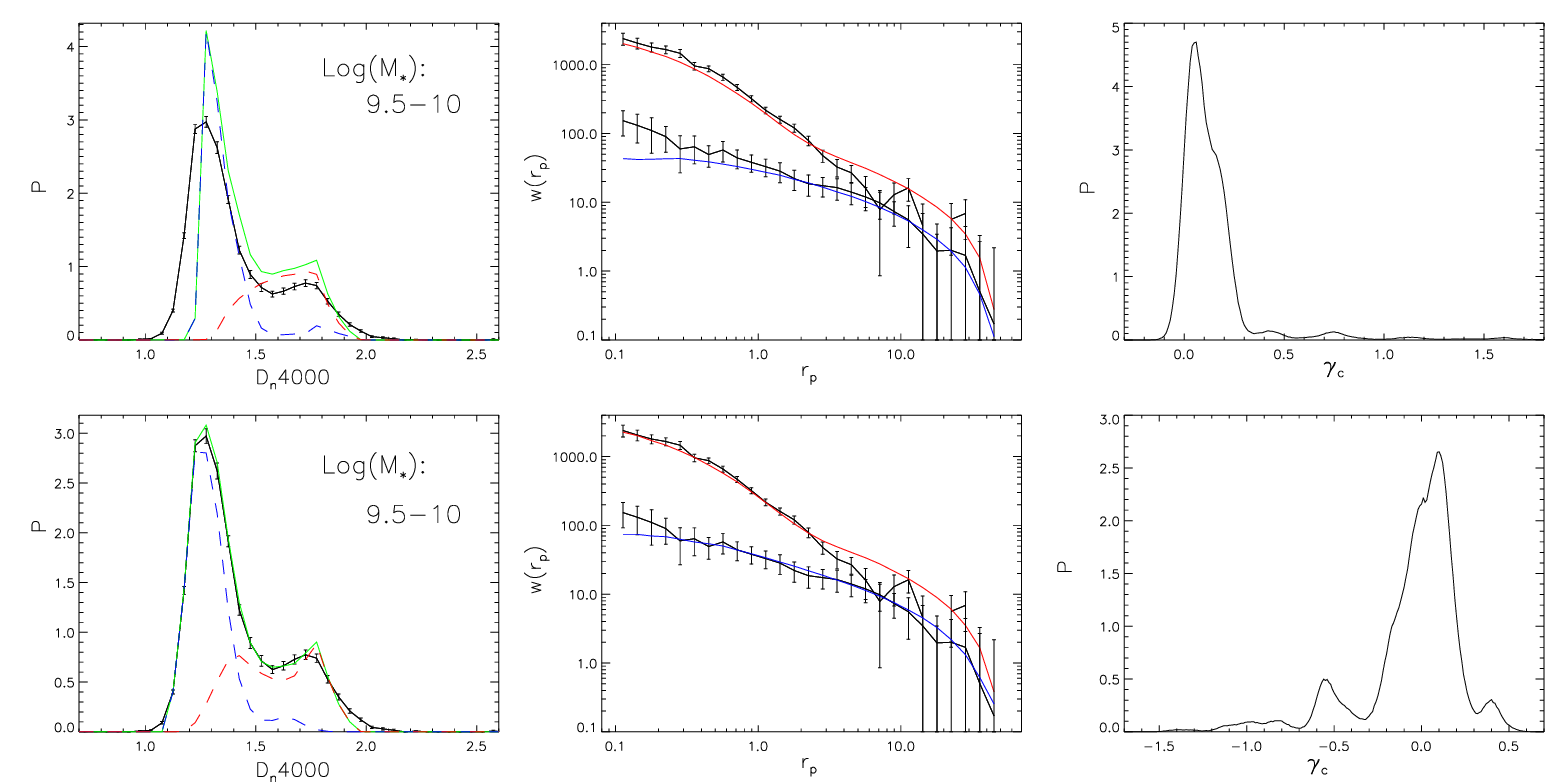}}\\%
\caption{  
Upper panels: the best fit when  $\gamma_c$ is constrained to take on
positive values.
Lower panels: the best fit when $\gamma_c$ is constrained to
lie below a value of 0.5.
Results are shown only for the lowest stellar mass bin:  
$10^{9.5}-10^{10}M_{\odot}$. 
}
\label{fig:nonegbestD4k}
\ec
\end{figure*}

\section{PARAMETRIZATION OF THE STAR FORMATION HISTORIES OF GALAXIES
IN THE MODEL}
\label{sec:fitting}

\subsection {Computation of the SEDs}

Recent semi-analytic models \citep{croton2005, bower2006, kang2006,
cattaneo2006b} have attempted to resolve some
of the difficulties outlined in the previous section by including
``AGN feedback''. The main effect of this form of feedback is to 
move the central galaxies of massive dark matter haloes to the red
sequence. 
Our approach is different; rather than {\em assume} some 
process that suppresses star formation in massive galaxies,
we {\em parametrize} the star formation histories of the galaxies
in our simulation using simple functions, and we allow the parameters
of our model to be constrained by the SDSS data.

The formation time of each galaxy is defined as the time when the halo 
of the galaxy is first found in the simulation. We also know   
the infall times of all satellite galaxies, i.e. the times
when they first became satellites. 
To model the star formation histories of our galaxies, we assume that the star 
formation rate of a galaxy declines  exponentially with time after its
formation and  depends on 
the stellar mass (at redshift $0$) of the galaxy. 
 We also assume that if a  galaxy is accreted and 
becomes a satellite, its star formation therafter declines with
a different e-folding time $\tau_s(M_*)$. The model  therefore 
has two timescales $\tau_{c}(M_{*})$ for central galaxies  
$\tau_{s}(M_{*})$ for satellite galaxies. 

The star formation rate of a galaxy can thus be written:

\[{SFR(t)}=\left\{\begin{array}{ll} 
e^{-t/\tau_{c}}& \mbox{, central galaxies}\\
e^{-t_{central}/\tau_{c}}e^{-(t-t_{central})/\tau_{s}}& 
\mbox{, satellite galaxies}\end{array} \right. \] 
where the  age of a galaxy $t$ is calculated starting from its formation time 
$t_{formation}$, which is assigned to a random time  between the 
simulation snapshot when the 
halo of the galaxy was  first found and the immediately preceding  snapshot.  
 $t_{central}=t_{infall}-t_{formation}$, is the 
time that the galaxy spends as the central object of
its own halo. 

The resulting  spectral energy distributions (SEDs) 
of the galaxies in the simulation are computed using the 
stellar population synthesis model of BC03 \citep{bruzual2003},
assuming a \citet{chabrier2003} IMF. 
Spectral properties such as the 4000 \AA\ break
strength and colour  depend 
on the metallicity of the galaxy as well as on
its star formation history. \citet{gallazzi2005} show that there exists a 
relation between stellar metallicity and stellar mass for galaxies
in the local Universe. We use the mean 
relation derived in their paper to specify the metallicity
of the galaxies in our simulation at a given value of $M_{*}$. 
Fig.~\ref{fig:BC03} shows the evolution of  
D$_n4000$/$g-r$ with time for three different values of the   
 $\gamma_c$ parameter ($\gamma_{c}$=1/$\tau_{c}(Gyr)$) for a typical central
galaxy in our model. Results are shown for solar metallicity
BC03 models (solid curves), as well as for 0.25 solar models (dashed curves).
As can be seen, there is a  small but significant dependence of
D$_n4000$ and $g-r$ on metallicity,
particularly for galaxies with short star formation time scales.  
Note also that this computation of the spectral energy distribution neglects
the effects of dust on the light emitted by the galaxy.
In the following sections,  we will concentrate on the 4000 \AA\
break strength, because of its very weak dependence on dust.

\begin{figure*}
\bc
\hspace{-1.6cm}
\resizebox{17.cm}{!}{\includegraphics{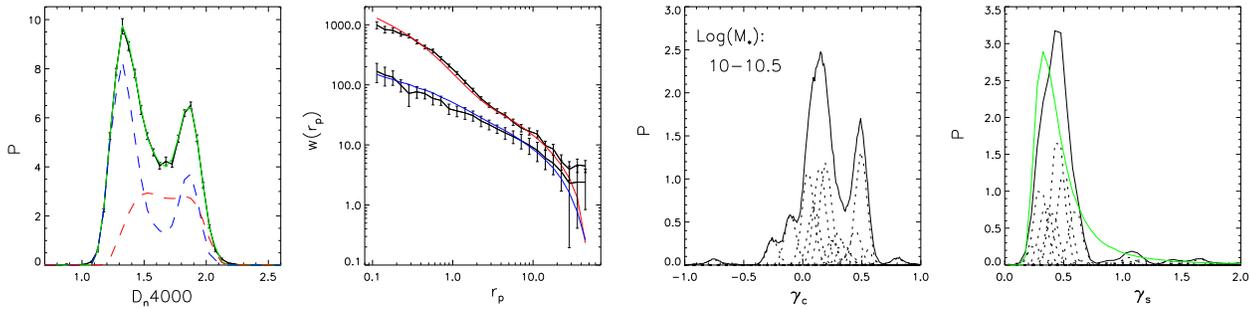}}\\%
\caption{ The results of a full non--parametric fit 
   to the distribution of star formation timescales of 
    satellite and central galaxies in the stellar mass range   
     $10^{10} -10.^{10.5} M_{\odot}$.
 In the right panel, the black curve shows the
 result of the non-parametric method, while the  green line is from
 the best-fit Gaussian of $\tau_s$.                                            
}
\label{fig:fullpar}
\ec
\end{figure*}

\subsection {Fitting the Data}

For each galaxy in the Millennium simulation, we have determined 
$t_{formation}$ and $t_{infall}$. 
We have also built a ``library''  of predicted present-day D$_n4000$ values by
running the BC03  models for different combinations of  
galaxy metallicity $z$, galaxy lifetime, and the two star formation time 
scales $\tau_{c}$ and $\tau_{s}$. We      
interpolate over the grid of  parameter values stored in the library to 
obtain D$_n4000$ for all the galaxies in the simulation. 
In addition, we know the (x,y,z) positions of the galaxies
within the simulation volume , and their  stellar masses
have already been specified using the model developed
as part of our previous  study \citep{wang2006}. We   therefore have all
the necessary ingredients to calculate the D$_n4000$ distributions as well 
as the correlation functions split by D$_n4000$ 
for different ranges in stellar mass and to compare the model
predictions  with the observations.

Initially we attempted to parametrize the distributions of $\tau_{c}$ 
and $\tau_{s}$
using simple Gaussian functions. However, this resulted in rather poor fits
to the data. Further experimentation indicated that it would be
advantageous to switch to a method that would allow the distributions
of $\tau_c$ and $\tau_s$ to take on complex shapes.
Following the methodology outlined by                                  
\citet{blanton2003a}, the distribution of $\gamma_{c}=1/\tau_{c}$
is parameterized by a sum of many Gaussians with mean values 
$\gamma_k$ equally distributed in a given range:
\begin{displaymath}
{P(\gamma_{c})}= \sum{N_{k}\frac{1}{\sqrt{2\pi\sigma^2}}e^{{-\frac{1}{2}(\frac{\gamma_{c}-\gamma_{k}}{\sigma})^2}}}{/}\sum{N_{k}}
\end{displaymath}

We assume the same scatter $\sigma$ for each Gaussian, but allow the  
weighting factors $N_k$  to vary.   
For each stellar mass interval, the range and number of Gaussians used in this 
non-parametric fitting technique are listed in Table 1. The central 
values $\gamma_k$ 
of each Gaussian are equally distributed over the range with a step of $0.05$.
$\sigma$ for each Gaussian is fixed to be the same as this step width.
This method allows the distribution of $\gamma$ values to take on any shape.
This approach turned out to be critical for central
galaxies, but resulted in little improvement when applied to the satellite 
population (see Sec. 5.1). We therefore employ the non-parametric
method only for the central galaxies. For satellite galaxies, a simple 
Gaussian centred at $<\tau_s>$ and with the scatter of $\sigma_{\tau_s}$ 
is used to parametrize the distribution of $\tau_s$.

During the fitting procedure, we also noticed that constraining 
$\gamma_{c}$ to be positive did not allow us  to reproduce
the blue end of D$_n4000$ distribution. We therefore 
allowed the parameter $\gamma_{c}$ to take on both  positive and  negative
values. In practice, a negative value of $\gamma_c$ corresponds to a galaxy
that is experiencing an elevated level of star formation at the present
day (i.e. a starburst).
For satellite galaxies, the time scale $\tau_c$ is constrained to be
positive, which is in any case preferred by our fits. 

\begin{figure*}
\bc
\hspace{-1.6cm}
\resizebox{17.cm}{!}{\includegraphics{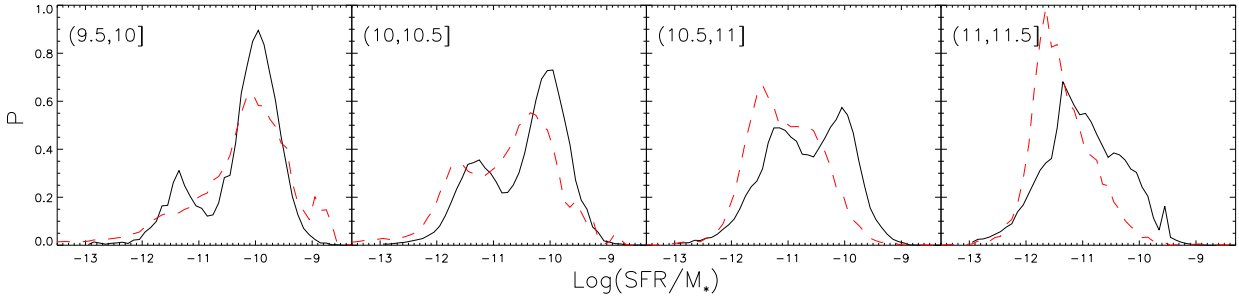}}\\%
\caption{  
The distribution of  specific star formation rates predicted by our model 
(red dashed lines) is compared with results derived from the SDSS 
(black solid lines), in different stellar mass bins.
}
\label{fig:SSFR}
\ec
\end{figure*}

\begin{figure*}
\bc
\hspace{-1.6cm}
\resizebox{17.cm}{!}{\includegraphics{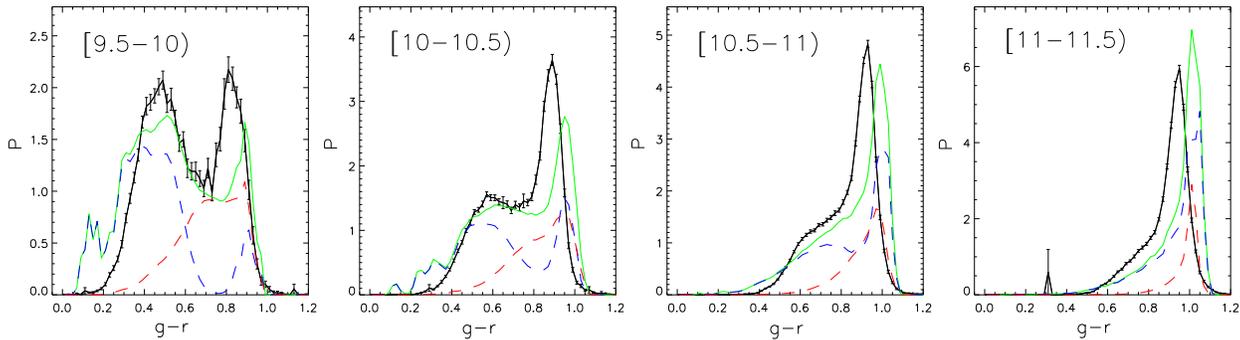}}\\%
\caption{  
$g-r$ distributions  predicted by our model compared to results from 
the SDSS. 
The green curves  show the full distribution,
while the red and blue dashed curves  show the contributions from
satellite and central galaxies.  
Black curves  show results from the SDSS.
}
\label{fig:bestgr}
\ec
\end{figure*}

To find a best-fit to both the D$_n4000$ distribution 
and the galaxy 
correlation functions split by D$_n4000$, we employ the Levenberg-Marquardt 
algorithm, which interpolates between the Gauss-Newton
algorithm and the method of gradient descent.
The final results of the fitting procedure are shown in 
Fig.~\ref{fig:bestD4k}. For each 
stellar mass bin, D$_n4000$ distributions and correlation functions 
for high/low D$_n4000$ subsamples are shown together with
the distribution of $\gamma_{c}$ that produces the best fit. 
In the left panel, blue and red dashed curves show results for
central and satellite galaxies, while  green lines are for both types of   
galaxy. In the middle panel, red and blue curves  
refer to high and low D$_n4000$ subsamples defined using
the same technique that was applied to  the SDSS
data. The distribution of 
$\gamma_{c}$ that results in these fits is shown  in the right panels.
Table 1 lists the parameters of the best fit models for four stellar
mass bins. For central galaxies, the median value of $\gamma_{c}$ is listed,
as well as the timescale $\tau_c=1/\gamma_c$, its median value and its 
16 and 84 percentile values.
The table also lists $<\tau_{s}>$ and  
$\sigma_{\tau_{s}}$, the parameters that describe
the star formation histories of satellite galaxies. 
We estimate our parameters by minimizing the quantity:
\begin{displaymath}
{\Xi}= \frac{{\chi}^2_{dis}}{N_{dis}}+\frac{{\chi}^2_{corr}}{N_{corr}}
\end{displaymath}
with
\begin{displaymath}
{{\chi}^2_{dis}}= \sum_{N_{dis}}{\left[\frac{P-P_{SDSS}}{\sigma(P_{SDSS})}\right]^2}
\end{displaymath}
and
\begin{displaymath}
{{\chi}^2_{corr}}= \sum_{N_{corr}}{\left[\frac{w(r_p)-w(r_p)_{SDSS}}{\sigma(w(r_p)_{SDSS})}\right]^2}
\end{displaymath}
where ${\chi}^2_{dis}$ is evaluated for
the D$_n4000$ distribution  and ${\chi}^2_{corr}$ is evaluated   
for the projected correlation functions of the high and the low  D$_n4000$
subsamples. For each stellar mass bin, $N_{dis}$
is the number of points along the  D$_n4000$ distribution that we fit.
In practice, we adopt $N_{dis}=50$ with D$_n4000$ in the range [0.5,3]. 
$N_{corr}$ is the number of points on the correlation function used in the 
fit. We adopt $N_{corr}=20\times2$
with  $r_p$ ranging from $0.113$ to $8.972h^{-1}$ Mpc.

\section{ THE RESULTS }
\label{sec:tests}

\subsection {Discussion of the Results}

From Fig.~\ref{fig:bestD4k} and Table.1, it is clear that star formation
histories of low mass and high  central galaxies are very different. 
At large stellar masses  
($M_{*}>10^{10.5}M_{\odot}$), nearly all central galaxies have positive 
$\gamma_c$, and a large fraction of them have 
ceased forming stars. There is a peak in the distribution of $\gamma_c$
at a value of around $0.45$, which corresponds to an e--folding timescale
of $\sim2$ Gyr.
Together with a minority of old and massive satellite galaxies,
these central galaxies in which  star formation has  shut down
are necessary to explain the strong peak in the D$_n4000$ distribution
at values of $\sim 1.8$,  characteristic  of  metal-rich, evolved stellar
populations.

\begin{figure*}
\bc
\hspace{-1.6cm}
\resizebox{17.cm}{!}{\includegraphics{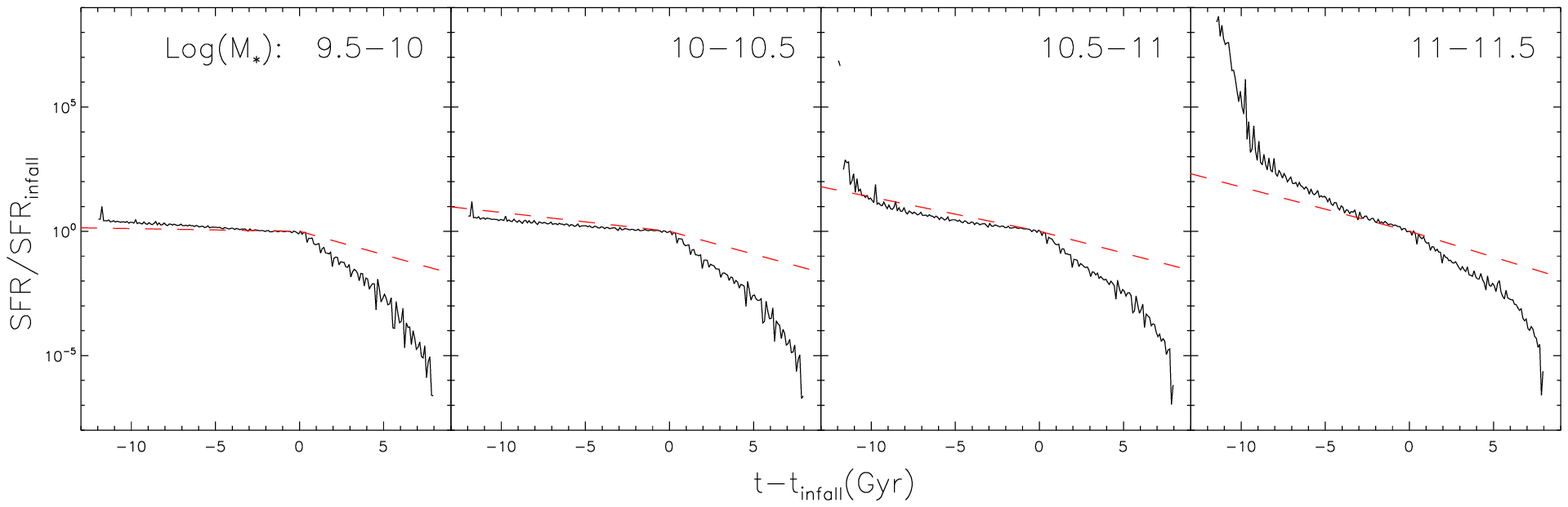}}\\%
\caption{
The median star formation histories of satellite galaxies in different 
stellar mass bins. Black lines show results from the 
semi-analytic models of \citet{lucia2006}, while
red dashed lines are from our models. The SFR is normalized to 
its value at the infall time. On the x-axis, the
time is plotted  relative to $t_{infall}$.
}
\label{fig:SFH}
\ec
\end{figure*}

At low stellar masses, galaxies display a much wider variety in                
star formation history. This is especially true in our  lowest stellar mass bin
($9.5<\log(M_{*}/M_{\odot})<10$). Nearly half the galaxies in this mass range
have  negative $\gamma_{c}$. In addition, 
there are also some objects that have ``switched off', i.e. the distribution
of $\gamma_c$ exhibits a tail toward large positive values.

In the upper panel of Fig.~\ref{fig:nonegbestD4k} we show
what happens to our fit in this stellar mass range if $\gamma_{c}$ 
is forced to take on only positive values. The fitting procedure
is exactly the same as described in Sec.4, 
but $\gamma_k$ is constrained to lie in the range $[0.05,2]$.  
The result shows that the 
blue end of the D$_n4000$ distribution is not well fit   
without a population of  ``star-bursting'' galaxies. 
In the BC03 model, a constant star formation 
rate will result in a  D$_n4000$ value of no less than $\sim1.2$ for 
a galaxy with  
an age of a few Gyrs (see green lines in Fig.~\ref{fig:BC03}).  
It is thus  not possible to obtain low enough values of D$_n4000$
to match the observations, unless we allow for 
negative values of  $\gamma_{c}$.

In the lower panel of Fig.~\ref{fig:nonegbestD4k},
we test what happens if we do not allow the distribution of
$\gamma_k$ to extend to very large positive values.    
If we truncate the distribution at a value of 0.5,
we still find fairly good fit with ${\Xi}=4.836$, 
which is comparable to our best-fit model.
This implies that the 
existence of a long tail of  galaxies with large
positive values of $\gamma_k$ is not strongly preferred, i.e. 
our model is not sensitive to the exact timescale over which the star
formation was truncated in these ``old'' central galaxies.  
One possibility is that these galaxies correspond to a post-starburst
phase in which  star formation has been temporarily reduced
following exhaustion of the gas  or blow-out of a
significant fraction of the interstellar medium.

From the values of $\tau_{s}$ listed in Table 1, it can be seen that 
unlike central galaxies,
satellite galaxies of all masses have similar e--folding timescales,
with an  average value of  around $2-2.5$ Gyr.
This indicates that all satellite galaxies experience a similar   
decline in star formation rate after falling into a 
large structure. To test the robustness of this
conclusion, and to see if a Gaussian is sufficient to describe the dispersion 
in $\tau_{s}$ values, we have carried out  a full non-parametric fit   
to constrain the shape of both the $\gamma_c$ and 
the $\gamma_s$($\gamma_s=1/\tau_{s}$) distributions for galaxies 
in the stellar mass range $10^{10-10.5}M_{\odot}$. The results are shown in
Fig.~\ref{fig:fullpar},  with the panel on the far right showing
the derived distribution of $\gamma_{s}$. The distribution of
$\gamma_{s}$ obtained using the non-parametric fitting method  
is somewhat different to the simple Gaussian fit (indicated on the plot
by the green curve). Nevertheless, the resulting median value of $\tau_{s}$
is $2.257$ Gyr, which is about the same as that of the simple Gaussian fit.
The small difference in the distribution of $\tau_s$ between 
the two methods makes very little difference to  
the fit or to the resulting $\Xi$ values.

\subsection {Consistency Checks}

So far, we have tuned the parameters $\tau_c$ and $\tau_s$  to reproduce
the D$_n4000$ distributions of SDSS galaxies as a function of stellar mass and
the projected correlation functions of high D$_n4000$ and low D$_n4000$ 
galaxy subsamples. We have chosen to focus on
D$_n4000$ because it is relatively insensitive to the effects
of dust. However,  D$_n4000$ is only one
of many possible age-indicators,  so it is interesting
to check whether we obtain consistent results for other measures
of the star formation history of a galaxy.

\citet{brinchmann2004} computed specific star formation rates (i.e. the
star formation rate per unit stellar mass) for SDSS galaxies.
For star-forming  galaxies, the 
values were  derived using emission line fluxes suitably corrected for
the effects of dust, so this indicator
of recent star formation history is independent of D$_n$4000.
For galaxies with absent or very weak emission lines, 
the specific star formation rates were in fact estimated using the
D$_n4000$ index,  so this is no longer an independent measure.                
Nevertheless it is interesting to test whether the {\em distribution}
of  SFR/$M_*$ predicted by our best-fit model is in agreement with
the results derived directly from the SDSS data.

In Fig.~\ref{fig:SSFR}, we show the results of this test.                   
The red dashed curves in the figure show the distributions of  the specific 
star formation rates predicted by our best fit model in four different
bins of stellar mass. 
To calculate these values, we simply integrate 
the star formation rate over the lifetime of a galaxy to get its total
stellar mass and we then divide the star formation rate
at the present day by this value. The fraction of stellar mass that is
returned to the interstellar medium over the lifetime of 
the galaxy is taken to be $0.5$, 
which is the median value predicted by the BC03 model.
The black lines show the results obtained directly from our SDSS sample,
which have beed corrected for volume incompleteness with the $1/V_{max}$
weighting scheme (the same method as for computing D$_n4000$ and $g-r$ 
distributions described in Sec. 2).  
From the plot we see that the agreement between our model ``predictions''
and the observations is reasonably good. The qualitative trends as a 
function of stellar
mass are reproduced quite well, but the absolute values of the specific
star formation rate predicted by the model tend to be somewhat lower
than in the data.  
  
We now test whether the same models allow us to reproduce the colour 
distributions
of galaxies. As we have noted, galaxy colours are quite sensitive to
the effects of dust, so it would be surprising to us if this
were the case! 

Fig.~\ref{fig:bestgr} shows the predicted $g-r$ colour distributions
in the absence of dust  and indeed,  it appears that the model does not fit
very well. At low stellar masses, the model produces too many
extremely blue galaxies and not enough galaxies in the red peak.
This tells us that a substantial number of galaxies in this red peak are
actually star-forming galaxies that are strongly reddened by dust.
At high stellar masses, the predicted shape of the $g-r$ colour distribution
is similar to the observations, but there is an offset in the predicted
colours of galaxies in the red peak as compared to
the observations. The existence of an  offset in $g-r$
of around $0.1$ mag  for high stellar mass galaxies has been noted 
several times in the literature on stellar population models
(see section 2.4.3 in \citet{gallazzi2005} and appendix of BC03
for more details about this colour mismatch for galaxies with
old stellar populations). One possibility is that the offset is related to the
effects of non-solar element abundance ratios on the SEDs, something
which is not currently accounted for by the BC03 models.

In summary, our results show that although very similar qualitative results
are obtained when using different spectral indicators, there are 
non-negligible quantitative differences. In particular, the effects of 
dust on galaxy colours must be taken into
account when interpreting  colour distributions and the clustering
properties of galaxies split by colour. 
This problem is not as serious for high mass
galaxies, which are mainly ellipticals with little gas, ongoing star formation
or dust. However, it is a major issue for low mass galaxies, where 
the shapes of 
the distribution of D$_n4000$ and $g-r$ are quite  different. 
As we have already seen from Fig.~\ref{fig:SDSS}, the relative fractions of
high/low D$_n$4000 and red/blue galaxies also differ  substantially 
in our lowest mass bins.

\subsection {Comparison with the Results of Semi-analytic Models}

As discussed in Sec. 1, the star formation histories of galaxies
predicted by modern semi-analytic models ought to be 
similar to those in our model. In the SAMs, the star formation rates
of  galaxies also decline after they transition from central
galaxies to satellite systems. In addition, modern SAMs also incorporate
AGN feedback mechanisms that act to shut down star formation in
the central galaxies of massive dark matter halos. It is interesting to
investigate whether there is ``quantitative'' agreement between our
results and those of the semi-analytic models.

In Fig.~\ref{fig:SFH}, we compare the median 
star formation history of satellite galaxies  in our model (red dashed lines) 
with the median star formation history
of satellites in the  semi-analytic galaxy catalogues  
of \cite{lucia2006}(black lines).  
Results are shown for four different stellar mass bins. 
On the x-axis, we plot the time relative to $t_{infall}$.
On the y-axis, we plot the star formation rate  normalized to its value 
at $t=t_{infall}$. As can be seen, the main difference between
the two models lies  in the behaviour of the star formation after the
galaxy is accreted as a satellite. In the semi-analytic model, star
formation declines much more rapidly than in our model. In our model,  
the median star formation
e-folding time scale of satellites is around $2-2.5$ Gyr, independent of galaxy
stellar mass. In the semi-analytic models, the 
e-folding time is closer to 1 Gyr. The median star formation rates
during the phase when the galaxies are central objects are  actually
quite similar in the two models, except for the very most massive galaxies.
In the semi-analytic models, most of the massive galaxies appear to have
experienced a short period of very intense  star formation in their past,
probably reflecting an early phase of gas-rich merging. 

\begin{figure*}
\bc
\hspace{-1.6cm}
\resizebox{12.cm}{!}{\includegraphics{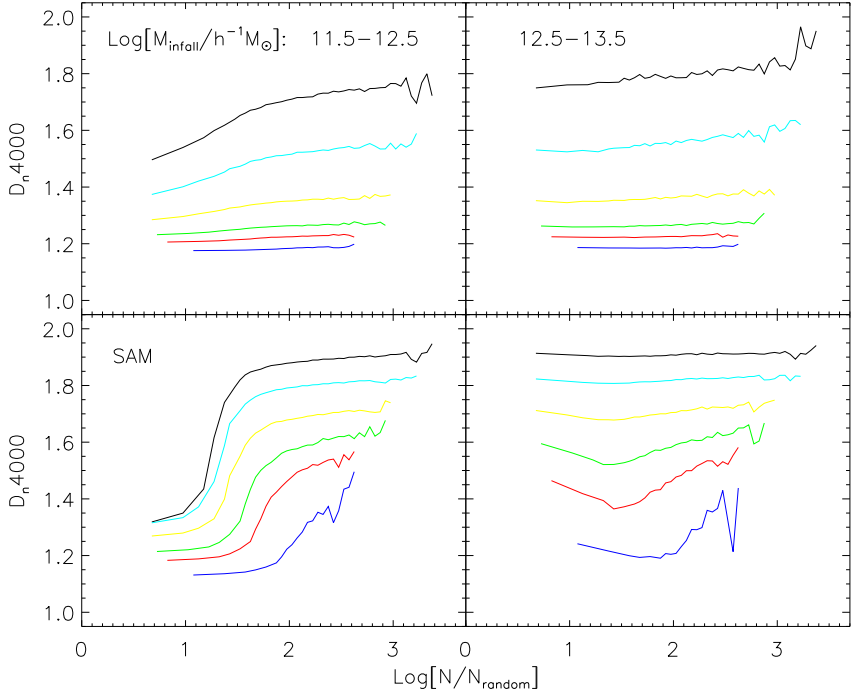}}\\%
\caption{
Colour--density relations for our  models (upper panels) and for the
semi-analytic
model(lower panels) in bins of hosting halo mass at the
infall time of the galaxy.
Coloured lines represent different redshifts: $z=0$, $0.3$, $0.8$, $1.5$, 
$2$, $3$ from top to bottom in each figure.
$N$ and $N_{random}$ refer to number of galaxies  within a sphere of 
radius $2Mpc$ centred on the galaxy  and on random positions. 
}
\label{fig:environ}
\ec
\end{figure*}

\subsection {Evolution to Higher Redshifts}

As we showed in the previous section,
there are significant  differences in the
star formation histories of satellite galaxies in our model as compared to
the semi-analytic model of \cite{lucia2006}.
As a result,   the colours and spectral
properties of satellites galaxies in the two models will    
differ substantially at high redshifts, because
the time between $t_{infall}$ and
$t_{observation}$ will be considerably smaller than at the
present day. If the star formation e-folding timescale
of satellite galaxies is long, one would expect many of
these systems to be still blue and actively star forming
at higher redshifts.  Recent work by  
\citet{cucciati2006} and \citet{cooper2006} has found that the steep 
colour--density relation observed at low redshift becomes weaker and 
gradually disappears by a  redshift of $\sim1.5$. 

In Fig.~\ref{fig:environ}, we plot the relations between D$_n4000$ 
and local density in our model (upper panels) and 
in the semi-analytic model (lower panels).
Results are shown at 5 different redshifts and for two different
ranges in the quantity $M_{infall}$. In the semi-analytic model, 
D$_n4000$ is calculated using the BC03 model{\footnote {Note that this index 
is not currently available in the publically released galaxy catalogues.}}.
The local density is evaluated by counting the number of galaxies
within a sphere of radius 2 Mpc centred on each galaxy and dividing by 
the corresponding number in randomly distributed spheres of the same size.
We only include galaxies with $M_{infall}$ greater than 
$2\times10^{11}h^{-1}M\odot$ when calculating these densities in order to
make sure that our results are not affected by the resolution
limit of the simulation.
Lines with  different colours show the
median $D_n4000$ for galaxies as a function of  density at 
redshifts $z=0$, $0.3$, $0.8$, $1.5$, $2$, $3$. In both models, the 
D$_n4000$--density relations are shallower for  galaxies with higher values
of $M_{infall}$, reflecting the fact that the star formation
histories for satellite galaxies are rather similar at high
masses.  However, the evolution with redshift
differs substantially in the two models. 
In our models, the D$_n4000$--density relations become much flatter at 
higher redshifts and they basically disappear by a redshift of $\sim1$.
In contrast, the relations in the semi-analytic models
remain steep up to redshifts $\sim3$.  
It is interesting that for galaxies
with  $M_{infall} > 
10^{12.5} M_{\odot}$, the trend in colour as a function of density
exhibits a non-monotonic behaviour at higher redshifts in the SAM. 

The main reason why the colour density relation of low mass galaxies  
remains  steep out to high redshifts in the semi-analytic models is because
$\tau_s$ is short. As we noted in the previous subsection, at redshift
zeros $\tau_s$ is $\sim 1$ Gyr in the SAMS as compared to $\sim 2.5$ Gyr for
our best-fit model. In SAMs, the star formation rate is assumed to be  
invesely proportional to the  dynamical time of the galaxy and hence, for
a given value of $M_{infall}$, $\tau_s$ will be smaller at higher
redshifts. In our model, we have assumed that $\tau_s$ is independent 
of redshift.

\section{Summary and Conclusions}
\label{sec:conclusion}

We have extended the physically-based halo occupation distribution (HOD) 
models of \citet{wang2006}, which relate the
stellar mass of the galaxy to the  mass of its halo  at the time it
was last a central dominant object.   
The models presented in this paper consider the  dependence 
of  clustering on the spectral
energy distributions (SEDs) of galaxies. The star formation
history of a galaxy is assumed to  depend on stellar mass and on the time
at which the galaxy transitions from being a   central galaxy to
a satellite galaxy orbiting within a larger structure.   
The stellar population synthesis 
model of \citet{bruzual2003} is used to compute the spectral energy 
distributions of the galaxies in the model.  
Rather than colour, we focus  on the spectral index D$_n4000$
because of its weak dependence on dust.
By fitting both the bimodal distribution of D$_n4000$  
and the projected correlation functions for high/low  
D$_n4000$ subsamples for SDSS galaxies in 4 different stellar
mass ranges, we constrain the star formation 
e--folding time of central  and satellite galaxies  as a function of stellar mass
at $z=0$.
 
Our results show that at high stellar masses, a large fraction of central 
galaxies have ceased forming stars. This shutdown in star formation is 
necessary to explain the fact that the majority of massive galaxies are red 
and old. It is also consistent with the conclusions of 
\citet{croton2005, bower2006,cattaneo2006b}, who show that a shut-down of 
gas cooling and star formation in massive dark matter halos results
in a much better fit to  present-day galaxy  luminosity functions
and colour distributions.
In these models, radio AGN feedback is invoked as the mechanism responsible for     
this suppression of cooling. 

For low stellar masses, central galaxies display a wide range of different 
star formation histories. A significant
fraction of low mass  central galaxies are experiencing starbursts.
We also find a ``tail'' of low mass  central galaxies in which star formation
is currently suppressed, but our model is not sensitive to 
exactly when this suppression occurred. 
In a recent paper, \citet{kauffmann2006} studied the  star formation histories
of local galaxies by analysing the scatter in their colours and spectral 
properties, and concluded  that star formation occurs 
in shorter, higher amplitude events in smaller galaxies.  It is interesting
that we come to very similar conclusions using a  completely different
method. Compared to central galaxies, satellites
have a much narrower distribution of star formation e--folding timescales. 
In satellites,  the average e-folding time  
does not depend on stellar mass and has  a value of around $2-2.5$ Gyr.

We have also checked whether our best--fit model parameters derived from
the distributions and clustering of galaxies as a function of  
D$_n4000$   can be extended to understanding the distribution of
galaxy colours.
Our conclusion is that the effects of dust must be taken into
account when modelling colours. This is especially true of low mass galaxies,
which can be  actively star-forming and yet  contain sufficient dust to   
match the colours of high mass ellipticals.

Finally, we have compared the star formation histories of galaxies in 
our model with the
 semi-analytic models of galaxy formation od De Lucia \& Blaizot (2007). 
The main difference between
the two approaches is in the timescale over which star formation declines
in satellite galaxies once they are accreted by larger structures.
In the semi-analytic models, star formation in satellites decreases  with  an
e-folding time of about $1$ Gyr. In our models, the decrease occurs
on a timescale that is a factor of $2-3$ times longer. This leads to a conflict between the
colours of satellites in the models and in the SDSS data. Similar conclusions
have been reached by \citet{weinmann2006}, who show that the semi-analytic
models of \citet{croton2005} predict a blue fraction of satellites that
is too low to match  results from their catalogue of galaxy
groups extracted from the SDSS.   

Assuming our derived star formation parameters to apply at
all times , we predict a weakening of the
colour--density relation towards higher redshifts and a complete
disappearance of this relation 
at redshift of about $1.5$. This is consistent with recent work    
by \citet{cucciati2006} and \citet{cooper2006}. In contrast,  
a strong colour--density relation is maintained in the
De Lucia \& Blaizot (2007)  semi-analytic model
up to redshifts greater than  $3$.

\section*{Acknowledgements}
We are grateful to Simon White for his detailed comments and 
suggestions on our paper. 
C.~L acknowledges the financial 
support of the exchange program between Chinese Academy of Sciences
and the Max Planck Society. G.~D.~L. thanks the Alexander von Humboldt 
Foundation, the Federal Ministry of Education and Research, and the 
Programme for Investment in the Future (ZIP) of
the German Government for financial support.

The simulation used in this paper was carried out as part of the programme of 
the Virgo Consortium on the Regatta supercomputer of the Computing Centre of 
the Max--Planck--Society in Garching. The halo data together with the galaxy data
from two semi-analytic galaxy formation models is publically available at
http://www.mpa-garching.mpg.de/milleannium/.

Funding for the SDSS and SDSS-II has been provided by the Alfred P. Sloan Foundation, 
the Participating Institutions, the National Science Foundation, 
the U.S. Department of Energy, the National Aeronautics and Space Administration, 
the Japanese Monbukagakusho, the Max Planck Society, and the 
Higher Education Funding Council for England. The SDSS Web Site is http://www.sdss.org/. The SDSS is managed by the Astrophysical Research Consortium 
for the Participating Institutions. The Participating Institutions are 
the American Museum of Natural History, Astrophysical Institute Potsdam, 
University of Basel, Cambridge University, 
Case Western Reserve University, University of Chicago, 
Drexel University, Fermilab, the Institute for Advanced Study, 
the Japan Participation Group, Johns Hopkins University, 
the Joint Institute for Nuclear Astrophysics, the 
Kavli Institute for Particle Astrophysics and Cosmology, the 
Korean Scientist Group, the Chinese Academy of Sciences (LAMOST), 
Los Alamos National Laboratory, 
the Max-Planck-Institute for Astronomy (MPIA), 
the Max-Planck-Institute for Astrophysics (MPA), 
New Mexico State University, Ohio State University, 
University of Pittsburgh, University of Portsmouth, 
Princeton University, the United States Naval Observatory, 
and the University of Washington.

\bsp
\label{lastpage}

\bibliographystyle{mn2e}
\bibliography{color}

\end{document}